\documentclass[aps,amsmath,amssymb,prl,twocolumn,superscriptaddress,showpacs]{revtex4}

\usepackage{bm}
\usepackage{epsfig}
\usepackage[usenames]{color}
\usepackage{graphicx}

\newcommand{\T}{{\cal T}}
\newcommand{\sT}{{\sf T}}

\newcommand{\av}[1]{\left\langle #1\right\rangle}
\newcommand{\bnabla}{\mbox{\boldmath$\nabla$}}

\newcommand{\Name}[1]{#1,}
\newcommand{\REVIEW}[4]{#1 {\bf #2}, #4 (#3).}

\newcommand{\avz}{{\overline z}}

\renewcommand{\paragraph}[1]{\textit{#1.---} }
\newcommand{\paper}{paper }

\newcommand{\etal}{{\it et al.}}

\begin{document}

\title{Giant Magnetoresistance in Nanogranular Magnets}
\author{A.~Glatz}
\affiliation{Materials Science Division, Argonne National Laboratory, Argonne, Illinois 60439, USA}
\author{I.~S.~Beloborodov}
\affiliation{Materials Science Division, Argonne National Laboratory, Argonne, Illinois 60439, USA}
\affiliation{James Franck Institute, University of Chicago, Chicago, Illinois 60637, USA}
\author{V.~M.~Vinokur}
\affiliation{Materials Science Division, Argonne National Laboratory, Argonne, Illinois 60439, USA}

\date{\today}

\begin{abstract}
We study the giant magnetoresistance of nanogranular magnets in the presence of an external magnetic field
and finite temperature. We show that the magnetization of
arrays of nanogranular magnets has hysteretic behaviour at low temperatures leading to a double peak in the magnetoresistance which coalesces at high temperatures into a single peak.
We numerically calculate the magnetization of  magnetic domains and the motion of domain walls in this system using a combined mean-field approach  and a model for an elastic membrane moving in a random medium, respectively. From the obtained results, we calculate the electric resistivity as a function of magnetic field and temperature. Our findings show excellent agreement with various experimental data.
\end{abstract}

\maketitle

\section{Introduction}

Recent experimental studies of magnetic
nanoarrays~\cite{zeng+prb06,ding+apl05,poddar+prb03,
kakazei+jap01,sankar+prb00,black+s00,xiao+prl92,
kakazei+tom99,schelp+prb97,rubin+epjb98,zhu+prb99, berkowitz+prl92,milner+prl96,levy-s92} revealed the giant magnetoresistance (MR) effect.
At low temperatures the hysteresis of the magnetization was found to lead to a splitting of the resistance peak~\cite{black+s00,xiao+prl92,kakazei+tom99,schelp+prb97,rubin+epjb98} whereas at temperatures above $\sim 100$K the magnetization hysteresis disappears and the MR shows a single peak at zero external magnetic field.
The interest in these systems is motivated both, by the
important technological promise~\cite{biosensors} and by the
opportunity of applying the ideas developed for granular materials for the description of doped manganite
systems~\cite{dagotto+prep01,manganites}.

In this \paper we investigate the giant MR of magnetic nanoarrays and develop a model for the field and temperature dependent magnetization hysteresis and the sample resistivity.  We consider temperatures below the Curie temperature of
the individual grains, $T_c^g$.
For $T<T_c^g$ the nanoarray can be either in a \textit{superferromagnetic} (SFM) state with the magnetic moments of the grains aligned by ferromagnetic interactions and the systems has a multidomain
structure~\cite{bedanta+prl07}, or, at higher temperatures, in a \textit{superparamagnetic} (SPM) state, where magnetic moments of individual grains (superspins) are mutually disoriented by thermal fluctuations.
In real systems the ferromagnetic coupling can appear due to a competition of exchange and dipole-dipole interactions.
The former is always present which favors ferromagnetic order, whereas the latter alone would lead to an antiferromagnetic order or a striped phase depending on the orientation of the (super-) spins.
Recently a detailed analysis of nano-granular arrays of magnetic particles revealed that short-ranged exchange interaction stabilize the superferromagnet and the ferromagentic interaction is provided by ''glue particles'', which are basically remains of the sample preparation process of nearly atomic size~\cite{bedanta+prl07}.

The electronic transport within a single domain in the
SFM state or in the SPM state is controlled by the average mutual alignment of the superspins~\cite{beloborodov+prl07} [see Eq. (\ref{eq.sigma}) below].
In addition, in the SFM state the contribution of the domain wall dynamics in an applied magnetic
field and the associated change of the size of domains with different bulk magnetizations on the MR has to be taken into account.

Magnetoresistance in granular materials was studied theoretically in Refs.~\cite{rubinstein-prb94,gu+prb96,pogorelov+prb98} before. However, these considerations were classical, using a spin-dependent scattering approach with a phenomenological spin-diffusion length, which is defined as the distance a spin-polarized conduction electron travels before it undergoes a spin-flip collision, for regular arrays of monodisperse spherical grains. In contrast to these previous considerations, we take Coulomb blockade effects, multiple electron co-tunneling, and domain wall dynamics into account in the present \paper. Moreover, our approach is applicable for arrays with position disorder, size distribution of the grains, quenched mesoscopic disorder due to impurities within the grains as well as in the embedding matrix. On top of the imperfections of the system, also annealed disorder in the form of thermal fluctuations of the magnetic moments is included.


\begin{figure}
\includegraphics[width=0.9\linewidth]{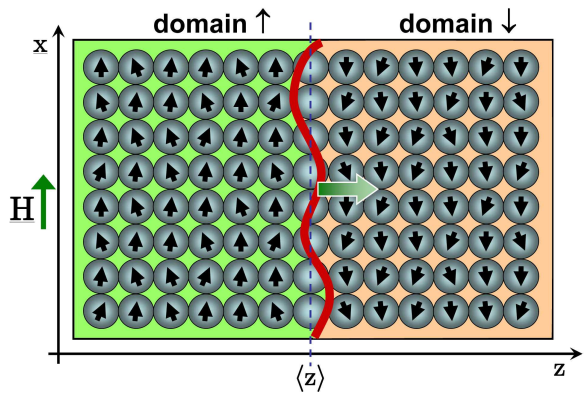}
\caption{Illustration of an idealistic array of ferromagnetic nano-particles with two magnetic {\it super-domains} in the SFM state. Our model also includes effects of quenched disorder induced by e.g. size and position fluctuations of the grains. The domain wall is considered to be thin compared to the domain size;  the bulk-magnetizations $m_{\uparrow ,\downarrow}$ of the two domains have opposite signs. The bold wavy line shall only illustrate the model of an elastic string which represents the real domain wall. In general domain walls are several grains thick where the average orientation of the superspins changes gradually from $\uparrow$ to $\downarrow$.
The dynamics of the domain wall is described by an elastic object driven by the external magnetic field $\mu_0{\bf H}$ moving in the $z$-direction.} \label{fig.model}
\end{figure}

\section{Electron transport}
We consider two- and three-dimensional arrays of conducting magnetic nano-size granules, weakly coupled via tunnel junctions.
Granularity gives rise to Coulomb blockade effects, controlling the low temperature conductivity at $T\ll E_c$ ($E_c=e^2/(\kappa a)$ is the charging energy of a single granule, with $e$ being the electron charge, $\kappa$ is the sample dielectric constant, and $a$ is the granule size. It can be as large as several hundred Kelvins which is exactly the experimentally relevant temperature interval)~\footnote{Due to the small grain size, electron energy is quantized with the mean level spacing $\delta$. Here we discuss the experimentally important case where $E_c/\delta \gg 1$}.
Then the conductivity of a nanogranular array composed of superspins in a {\it single} super-domain in the SFM (or in the SPM) state is given by~\cite{beloborodov+prl07}
\begin{equation}\label{eq.sigma}
\sigma (T,m^2) \sim g_t^0(1 + \Pi^2 m^2) \exp(-\sqrt{\T_0(m^2)/T}),
\end{equation}
where $g_t^0$ is the tunneling conductance of the SPM state with no magnetic field, $\Pi$ the polarization factor of the grains, $m^2=\av{\cos\theta}$ the normalized bulk magnetization ($\theta$ is the angle between the magnetic moments of adjacent superspins and $\av{\ldots}$ stand for averaging over a single domain), and $\T_0(m^2)\equiv T_0[1-(\xi_0/a)\ln(1+\Pi^2 m^2)]$ with $T_0=e^2/(\kappa\xi_0)$ and $\xi_0$ being the characteristic energy scale and the inelastic localization length, respectively.
Equation~(\ref{eq.sigma}) describes the contribution from the
co-tunneling processes governing hopping conductivity in granular materials~\cite{beloborodov+rmp07} which holds as long as the electron hopping distance exceeds the size of a single grain, $a$, i.e. as long as $T\lesssim T_0$.


The MR in nanogranular magnets (see Fig.~\ref{fig.model}) stems from (i) the field dependent bulk magnetization and (ii) the domain wall motion in the SFM state.
Normally the thickness of the domain walls is negligible as compared to the domain size itself, allowing to model their dynamics under the external magnetic field by an elastic membrane driven through a random medium~\cite{ioffe87,blatter+rmp94,brazovskii+ap04,glatz+prl03,petracic+prb04}.
We also exclude domain nucleation processes within a domain
involved and assume that the domain walls are sufficiently spatially separated or the external magnetic field changes fast enough in order to avoid effects of domain wall interaction (for more details, see discussion at the end of the \paper).
Thus, only the change of domain sizes associated with the domain walls dynamics contribute to the electron transport but not to the electron scattering at the walls~\footnote{The contribution of the domain wall to the conductivity can be estimated as follows:
As can be seen from Eq.~(\ref{eq.sigma}) the conductivity $\sigma(T,m^2)$ can decrease by about a factor of $2$ if $m$ changes from $1$ (in a domain) to $0$. This is approximately the case at a domain wall (in the worst case): Inside the domain we can assume that $m^2=1$, i.e. it has the best conductivity, and at the domain wall the average change of angles $\theta$ is at most $\pi/2$ if the wall contains at least one grain (a direct jump from the domain $\uparrow$ to the domain $\downarrow$ is unphysical).
The highest possible resistivity within the wall is therefore $1/\sigma(T,0)$ from which we can estimate the linear resistance of a singe domain wall of width $w$ as $R_{dw}\leq w/\sigma(T,0)\leq 2w/\sigma(T,m^2)$. Since we assume that the average domain size $L_{dom}$ is much larger than $w$ and the linear resistance of a domain is $R_{dom}\approx L_{dom}/\sigma(T,m^2)$ it follows immediately that the contribution of the domain wall to the total system resistance can be neglected. This assumption is in good agreement with the LMOKE [Longitudinal Magneto Optical Kerr Effect] micrographs shown e.g. in Ref.~\cite{bedanta+prl07}, where no effect of domain walls is visible.}. Incorporating the field dependent bulk magnetization and domain dynamics into the description of electron transport of Ref.~\cite{beloborodov+prl07}, we construct a theory of the giant magnetoresistance in magnetic nanoarrays.


\section{Magnetodynamics}
First, we discuss the bulk magnetization in the SPM state and/or in a single domain in the SFM state, using the mean-field (MF) equation~\cite{helman+prl76}
\begin{equation}\label{eq.MF}
m(h,\sT)=L(\alpha)\,,\,\,\text{with}\,\, \alpha=[\eta h+3L(\alpha)]/\sT\,,
\end{equation}
where $L(\alpha)=\coth(\alpha)-1/\alpha$ is the {\it Langevin}
function, and we introduce the following dimensionless parameters:
$h=H/H_0$, $\sT=T/T_c^s$, and $\eta=\mu H_0/(k_B T_c^s)$. Here $H_0$ is a characteristic magnetic field strength and $\mu$ is a material dependent density of magnetic moments.
In experiment~\cite{black+s00} the typical values are $\mu_0 H_0\approx 0.2$T, $T_c^s\approx 100$K, and $\eta\approx 1$.
Equation~(\ref{eq.MF}) is valid in the SPM state,  and in the SFM state for $h>h_c(\sT)$ with $h_c(\sT)<0$ being the critical field where the MF approach breaks down.
At this field $\alpha(h)$ in Eq.~(\ref{eq.MF}) becomes discontinuous.

Next, we turn to the SFM state. The equation of motion for an
over-damped $D$--dimensional interface profile $z({\bf x},t)$,
describing the domain-wall, embedded in a $d=D+1$ dimensional system has the following form~\footnote{We neglected an inertial term $\propto\frac{\partial^2 z}{\partial t^2}$ which is justified since the motion of the domain wall is assumed to be overdamped~\cite{brazovskii+ap04}.}~\cite{feigelman-jetp83}:
\begin{equation}\label{eq.motion}
\frac{1}{\gamma}\frac{\partial z}{\partial t}=\Gamma\bnabla^2z+
f(t)+g({\bf x},z)+\eta({\bf x},t)\,.
\end{equation}
Here $\gamma$ and $\Gamma$ denote the mobility and the stiffness constant of the interface, respectively, and $f(t)$ is the time dependant driving force ($f(t) = 2\mu_0\mu_B H(t)$ with $\mu_B$ and $\mu_0 H(t)$ being the Bohr magneton and magnetic field, respectively).
Random forces $g({\bf x},z)$ are defined by
$\av{g({\bf x},z)}=0$ and $\av{g({\bf x},z)g({\bf
x}^{\prime},z^{\prime})}= \delta^D({\bf x}-{\bf
x}^{\prime})\Delta_0(|z-z^{\prime}|)$. We further assume that
$\Delta_0(|z|)$ is a monotonically decreasing function of $z$ for $z>0$ which decays to zero over a finite distance
$\ell$,~\footnote{$\Delta_0(|z|)$ is typically a Gaussian function with variance $\ell$ which is the maximum of the domain wall width and the random force variance.}.
Thermal fluctuations are described by the random noise term $\eta({\bf x}, t)$ with $\av{\eta ({\bf x},t)}=0$ and $\av{\eta({\bf x},t)\,\eta({\bf x}^{\prime},t^{\prime})}= 2\frac{T}{\gamma}\,\delta^D({\bf x}-{\bf x}^{\prime})\,\delta(t-t^{\prime})$.
For the details of model~(\ref{eq.motion}) we refer the reader to Ref.~\cite{brazovskii+ap04}.

If the system is driven adiabatically, i.e. if it is always in a steady state even if the driving force is changing, the mean
velocity $v(t)\equiv \av{\partial z({\bf x},t)/\partial t}_{\bf x}$~\footnote{Here $\av{\ldots}_{\bf x}$ denotes the average over coordinates ${\bf x}$ (and disorder, which is already included since the system is self-averaging). Otherwise the average has to be taken over all function parameters.} shows a depinning transition at zero temperature or creep at finite temperatures~\cite{blatter+rmp94}.
As we are interested in the finite temperature regime and aim to describe a finite system, we restrict the domain wall by $|z({\bf x},t)|\leq L_z$, and define the magnetization parallel to the
magnetic field as
\begin{subequations}
\begin{equation}
M_{\|}(\sT,h) = \frac{M_s}{2}\left[(1+ \avz )m_{\uparrow} +
(1- \avz)m_{\downarrow}\right]\label{eq.m}\,,
\end{equation}
where $M_s$ is the saturation magnetization,
$\avz=\avz(T,h)\equiv \av{z({\bf x},t)}_{\bf x}/L_z$
the dimensionless mean displacement of the domain wall, and $m_{\uparrow,\downarrow}=m_{\uparrow,\downarrow}(\sT,h)$ the bulk magnetization for the two domains with opposite mean direction with $m_{\uparrow}(\sT,h)=-m_{\downarrow}(\sT,-h)$. The normalized sample magnetization is therefore \begin{equation}\label{eq.ms}
m_s(\sT,h)\equiv M_{\|}(\sT,h)/M_s\,,
\end{equation}
\end{subequations}
i.e. $m_s(\sT,h)\in [-1,1]$.

\begin{figure}
\includegraphics[width=0.8\linewidth]{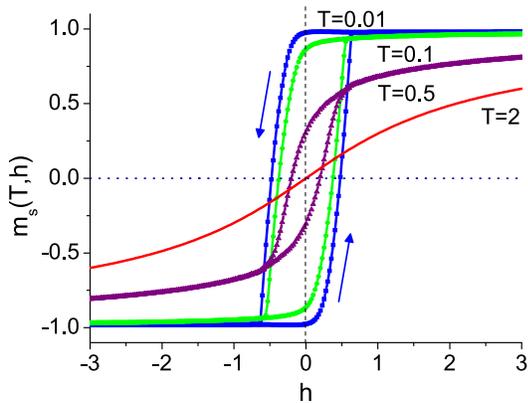}
\caption{Magnetization hysteresis, Eq. (\ref{eq.ms}), for different temperatures below $\sT=T/T_c^s<1$ and above the Curie temperature of the sample according to Eq.~(\ref{eq.MF}).  The arrows show the direction of hysteresis with changing magnetic field. The dimensionless magnetic field $h$ is given in units of $\mu_0 H_0$.} \label{fig.hyst}
\end{figure}

\section{Simulations}
We solve Eq.~(\ref{eq.motion}) numerically for various temperatures and for $f(t)$ slowly changing between $-f_{\max}$ and $f_{\max}$, such that non-adiabatic effects like the hysteresis of the domain-wall velocity $v(t)$ can be
neglected~\cite{glatz+prl03,petracic+prb04}. We introduce
dimensionless coordinates and time in the following way: The units of (simulation) time $\tau_0$ and space $\lambda_0$ are chosen such that $\gamma\Gamma\tau_0=\lambda_0^2$ and that the dimensionless random pinning forces $\tilde g=\tau_0\gamma/\lambda_0 g$ are uniformly distributed in $[-1/2,1/2]$.
The dimensionless driving force is $h\equiv f\lambda_0/(\tau_0\gamma)$ and the dimensionless temperature $\tilde \sT=\tau_0\gamma/\lambda_0^{2+D} T$.

We now discretize Eq.~(\ref{eq.motion}) in ${\bf x}$-directions in $N^D$ lattice sites (for $D=1$ the {\it lattice Laplacian} is given by $\bnabla^2 z_i=z_{i+1}+z_{i-1}-2z_i$, $i=1,\ldots,N$.). The stochastic forces $g({\bf x}, z)$ are constructed by choosing a random number in $[-1/2,1/2]$ at $z$-positions apart by $\ell$. In between the forces are linearly interpolated which results in a Gaussian $g$-$g$ correlator with variance~$\ell$,~\cite{glatz-phd}.

The dimensionless sample magnetization $m_s(\sT,h)$ for different temperatures $\sT$ and $h_{\max}=3$ is shown in Fig.~\ref{fig.hyst}.
By comparison with experimental data of Ref.~\cite{black+s00} the dimensionless simulation temperature $\tilde\sT$ corresponds to $\sim \sT/10$ and $h=1$ to a magnetic field of $0.2$T.
Notice, that for $|h(t)|>h_c(\sT)\approx 2.5$ at $\sT = 0.01$ the contribution of the opposite sign domain with respect to $h$ can be neglected for the sample magnetization.
Since in real experiments the external magnetic field $h$ cannot be changed adiabatically/infinitly slow, the driving force in the simulation is also changed very slowly but still oscillating, resulting in hysteretic behaviour of $m_s(\sT,h)$ even at finite temperatures.

\begin{figure}
\includegraphics[width=0.8\linewidth]{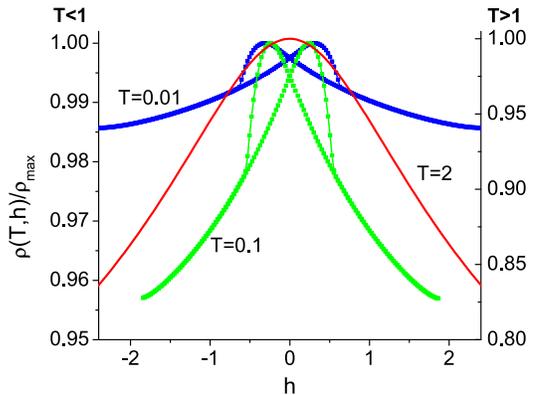}
\caption{Magnetic field dependent magnetoresistance for various temperatures. Temperatures are given in units of $T_c^s$. At low temperatures $\sT<1$ the magnetization hysteresis leads to the double peak structure of the MR (left y-axis). In the SPM regime ($\sT>1$) the double peak is coalesced into a single more pronounced peak (right y-axis). The resistivity is normalized to the maximal resistivity.} \label{fig.MR}
\end{figure}

Using the results for the bulk magnetization and the domain wall dynamics below $T_c^s$ we can now calculate the magnetoresistance for our model. In the SFM regime we have to distinguish the two cases of  parallel and perpendicular resistivity with respect to the domain wall:
\begin{subequations}
\begin{eqnarray}
\rho_{\perp}(\sT,h) &=& \frac{1}{2}\left[ \frac{1+ \avz}{\sigma(\sT,m_{\uparrow}^2)} +
\frac{1 - \avz}{\sigma(\sT, m_{\downarrow}^2)}\right], \\
\rho_{\|}(\sT,h)&=& \frac{2}{(1+\avz) \sigma(\sT,m_{\uparrow}^2) + (1-\avz) \sigma(\sT,m_{\downarrow}^2)}\,.
\end{eqnarray}
\end{subequations}
We note that $\rho_{\perp}(\sT,0)=\rho_{\|}(\sT,0)$ since $m_{\uparrow}^2(\sT,0)=m_{\downarrow}^2(\sT,0)$.
In the SPM state the domain structure does not exist. Thus, the electron transport is governed by the behaviour of bulk magnetization described by Eq.~(\ref{eq.MF}).

Using typical values for the parameters $\Pi^2 = 0.3$, $\xi_0/a = 1$, $T_0/T_c^s = 10$, we can calculate the magnetic field dependent magnetoresistance, $\rho(\sT,h)= [\rho_{\perp}(\sT,h)+\rho_{\|}(\sT,h)]/2$ in the SFM state and $\rho(T,h)=1/\sigma(T,m^2)$ in the SPM state where $m(T,h)$ is derived from Eq.~(\ref{eq.MF}).
The resulting curves for various temperatures are shown in Fig.~\ref{fig.MR}.
At low temperatures, $\sT<1$, a clear double peak in the resistivity of the system can be seen which is a direct result of the magnetization hysteresis.
The observed peak depends on the sign of the external magnetic field change, i.e. the right (left) peak for increasing (decreasing) field. In the SFM state the resistivity can only be calculated for $|h|<h_c(\sT)$, where the bulk magnetization varies only slightly such that there is no visible difference between $\rho_{\perp}$ and $\rho_{\|}$. For $\sT>1$ the MF approach holds for all $h$.

The maximum change of the resistivity with magnetic field can be quantified using the expression $\Delta \rho(\sT)\equiv(\rho_{\max}(\sT)-\rho_{\infty}(\sT))/\rho_{\max}(\sT)$, where $\rho_{\max}(\sT)$ is the maximum (peak) resistivity and $\rho_{\infty}(\sT)$ the ''saturation'' resistivity for the fully magnetized state [$m^2(\sT,h\to\pm\infty)=1$]. The factor by which $\rho$ changes is given by $\gamma=\Delta\rho(\sT)/[1-\Delta\rho(\sT)]$.

In the SFM state we get $\Delta\rho(\sT=0.01)\approx 0.032$ or $\gamma=3\%$ and $\Delta\rho(\sT=0.1)\approx 0.107$ or $\gamma=12\%$. The decrease of the effect towards lower temperatures is due to the fact that the bulk magnetization is less field dependent.
In the SPM state we can write an explicit expression for $\Delta \rho(\sT)$, using Eq.~(\ref{eq.sigma}), given by
$\Delta\rho(\sT)=1-(1+\Pi^2)^{-1}e^{-\sqrt{T_0/T}+\sqrt{\T_0(1)/T}}$.
For $\sT=2$ we get $\gamma=78\%$. Note, that within the field range of Fig.~\ref{fig.MR} the curve for $\sT=2$  does not saturate.

We now compare our results with reported experimental data.
In Refs.~\cite{zeng+prb06,ding+apl05,poddar+prb03,kakazei+jap01,sankar+prb00,black+s00,xiao+prl92}, the magnetization curves for magnetic nanoarrays were measured mostly at high temperatures -- showing a weak hysteresis effect.
In Ref.~\cite{black+s00} the magnetization data for weakly coupled $10$-nm $Co$-nanocrystals has been obtained at $5$K where the magnetization hysteresis has a width of approximately $0.1$T leading to the double peak structure of the resistivity. The resulting maximal resistivity change $\gamma_{\rm exp}$ is about $6\%$. The double peak structure in the resistivity was observed in several experiments~\cite{black+s00,kakazei+tom99,schelp+prb97,xiao+prl92}, with $\Delta\rho$ in the range $2\%$ to $10\%$, which is in good agreement with our results.
However, mostly the single peak in the giant magnetorestance in the SPM state at high temperatures (from $100$K to room temperature) has been studied~\cite{zeng+prb06,kakazei+jap01,sankar+prb00,kakazei+tom99,zhu+prb99,schelp+prb97,berkowitz+prl92}.
It would be interesting to compare our predictions for the magnetoresistance $\rho(\sT,h)$ to more precise measurements at low temperatures near the sample Curie temperature.

\section{Discussion}
There are several characteristic length scales in the present problem. Some of them are relevant for the description of the magnetization $m_s^2(\sT,h)$ while others for the electron conductivity $\sigma(\sT,m^2)$. To describe the magnetization behaviour in the SFM state we used the elastic model for domain walls driven in a random environment, Eq.~(\ref{eq.motion}): comparing the elastic (first) and random force (third) terms in  the r.h.s. of Eq.~(\ref{eq.motion}) one can show that the Imry-Ma or Larkin length scale is given by $L_p\approx [(\ell \Gamma)^2/\Delta_0(0)]^{1/(4-D)}$. On this scale the weak random forces accumulate to a value comparable to the elastic force. On smaller length scales the domain-wall is essentially flat and above the interface becomes rough. Therefore the domain size $L_{dom}$ in nanogranular magnets should be larger than $L_p$ in order to use the model for domain wall dynamics. Thus, our consideration is valid for the following sequence of length scales, $L_{\text{dom}} > L_p > a$. On the other hand the relevant length scale for electron conductivity is the hopping length $r_{\text{hop}}(T) \sim a \sqrt{T_0/T}$. Since the energy scale $T_0 \sim E_c > T$ the hopping length cannot be smaller than the size of a single grain, $r_{\text{hop}}(T) > a$. At the same time the relation between $r_{\text{hop}}(T)$ and $L_{\text{dom}}$ is arbitrary.

The mapping of simulation temperature and magnetic field slightly depends on the rate of change of the external driving force $h(t)$ in Eq.~(\ref{eq.motion}) - if it is changed more slowly, hysteresis effects will become smaller. If the external driving force $h(t)$ is changed faster and the domain wall motion becomes non-adiabatic - in  the sense that e.g. the velocity hysteresis becomes pronounced~\cite{glatz+prl03} -- several effects should be mentioned~\cite{petracic+prb04,kleemann-armr07}: First the magnetization hysteresis gets wider since the domain wall cannot follow the changes of the magnetic field anymore, leading to a greater separation of the double peaks of $\rho(T,h)$. Furthermore the change in the magnetization from $m=\pm 1$ to $m=\mp 1$ will become smeared out leading to a broadening of the resistivity peaks. In addition to these {\it visual} effects a phase shift between the magnetic field and magnetization appears.
The shape of the magnetization curves near the saturation points depends also on boundary effects - in the simulation $z$ was simply restricted by the system border, but no additional surface pinning effects were taken into account.
The qualitative picture remains, however, the same.

In our model we consider only one domain wall. Therefore, we comment under which conditions this model is justified, in particular why dipolar domain wall interaction can be neglected. Since the dipolar interaction between walls is of Coulomb type it is proportional to $1/\ell^2$ (in 2D), where $\ell$ is the typical distance between domain walls~\cite{haldane+jp81}.
From this we can introduce a typical time scale for the domain wall interaction as $\tau =\hbar \ell/q^2$, where $q$ is the charge of the walls. If the frequency of the external magnetic field is larger than $1/\tau$ the contribution of domain wall interaction can be neglected.
In other words if domain wall fluctuations due to interaction are slow (e.g. for large spatial distances between them) compared to the motion induced by the driving force, the former motion has no influence on the sample resistivity.
We assume this in our model and therefore consider only one domain wall.
The spiked domain structure observed in LMOKE micrographs in Ref.~\cite{bedanta+prl07} cannot be described in detail by our model. However, the effect of these structures on the magnetoresistance can be assumed to be small such that a coarse grained picture with only one wall will describe the transport correctly, at least on a  qualitative level.
We remark, that samples with a single domain wall can be created, see e.g. Ref.~\cite{lemerle+prl98} where an ultrathin Pt/Co/Pt film was prepared with one moving wall, and model (\ref{eq.motion}) was successfully applied to describe the magnetic susceptibility of nanogranular magnets~\cite{petracic+prb04}.

In general, also dipolar interactions within one domain wall have to be considered in the Hamiltonian for the wall, leading to an additional term in Eq.~(\ref{eq.motion}). This term describes the interaction between dipoles via Coulomb forces. Since the dipolar axis is usually parallel to the wall this additional term represents only a renormalization of the elastic term which can be seen in Eq.~(25) of Ref.~\cite{nattermann+jpc83}.
Therefore its influence is hidden in our phenomenological constant $\Gamma$ in Eq.~(\ref{eq.motion}).

In conclusion, we have studied the giant magnetoresistance of nanogranular magnets in the presence of an external magnetic field and finite temperature. We have shown that the magnetization of
arrays of nanogranular magnets has hysteretic behaviour at low temperatures leading to a double peak in the magnetoresistance which coalesces in the SPM state to a single one.
We have numerically calculated the magnetization of this system based on the model for domain walls moving in a random medium combined with a MF approach for the bulk magnetization, Fig.~\ref{fig.hyst}. Using the simulation results for the magnetization we have calculate the electric resistivity $\rho(T,h)$ as a function of magnetic field and temperature in the Ohmic regime, Fig.~\ref{fig.MR}.  Our findings are in a good agreement with published experimental data~\cite{sankar+prb00,black+s00,kakazei+tom99,schelp+prb97}.

\acknowledgments
We thank Wai-Kwong Kwok and Alexei Snezhko for useful
discussions. This work was supported by the U.S. Department of
Energy Office of Science through contract No. DE-AC02-06CH11357.
A.~G. acknowledges support by the DFG through a research grant.
I.~S.~B. was supported by the UC-ANL Consortium for Nanoscience
research.


\begin{thebibliography}{0}

\bibitem{zeng+prb06}
  \Name{H. Zeng \etal}
  \REVIEW{Phys. Rev. B}{73}{2006}{020402(R)}

\bibitem{ding+apl05}
  \Name{Y. Ding and S.~A. Majetich}
  \REVIEW{Appl. Phys. Lett.}{87}{2005}{022508}

\bibitem{poddar+prb03}
  \Name{P. Poddar \etal}
  \REVIEW{Phys. Rev. B}{68}{2001}{214409}

\bibitem{kakazei+jap01}
  \Name{G.~N. Kakazei \etal}
  \REVIEW{J. Appl. Phys.}{90}{2001}{4044}

\bibitem{sankar+prb00}
  \Name{S. Sankar \etal}
  \REVIEW{Phys. Rev. B}{62}{2000}{14273}

\bibitem{black+s00}
  \Name{C.~T. Black \etal}
  \REVIEW{Science}{290}{2000}{1131}

\bibitem{xiao+prl92}
  \Name{J.~Q. Xiao, J.~S. Jiang, and C.~L. Chien}
  \REVIEW{Phys. Rev. Lett.}{68}{1992}{3749}

\bibitem{kakazei+tom99}
  \Name{G.~N. Kakazei \etal}
  \REVIEW{IEEE Trans. on Magn.}{35}{1999}{2895}

\bibitem{schelp+prb97}
  \Name{L.~F. Schelp \etal}
  \REVIEW{Phys. Rev. B}{56}{1997}{R5747}

\bibitem{rubin+epjb98}
   \Name{S. Rubin, M. Holdenried \and H. Micklitz}
   \REVIEW{Eur. Phys. J. B}{5}{1998}{23}

\bibitem{zhu+prb99}
  \Name{T. Zhu \and Y.~J. Wang}
  \REVIEW{Phys. Rev. B}{60}{1999}{11918}

\bibitem{berkowitz+prl92}
 \Name{A.~E. Berkowitz \etal}
  \REVIEW{Phys. Rev. Lett.}{68}{1992}{3745}

\bibitem{milner+prl96}
  \Name{A. Milner, A. Gerber, B. Groisman, M. Karpovsky, and A. Gladkikh}
  \REVIEW{Phys. Rev. Lett.}{76}{1996}{475}

\bibitem{levy-s92}
  \Name{P.~M. Levy}
  \REVIEW{Science}{256}{1992}{972}

\bibitem{biosensors}
  \Name{D.~L. Graham \etal}
  \REVIEW{TRENDS in Biotechnology}{22}{2004}{455}

\bibitem{dagotto+prep01}
  \Name{E. Dagotto \etal}
  \REVIEW{Phys. Rep.}{344}{2001}{1}

\bibitem{manganites}
  \Name{Y.~D.~Chuang \etal}
  \REVIEW{Science}{292}{2001}{1509};
  \Name{Y.~Moritomo \etal}
  \REVIEW{Nature}{380}{1996}{141}

\bibitem{bedanta+prl07}
  \Name{S. Bedanta \etal}
  \REVIEW{Phys. Rev. Lett.}{98}{2007}{176601}

\bibitem{beloborodov+prl07}
  \Name{I.~S. Beloborodov, A. Glatz, and V.~M. Vinokur}
  \REVIEW{Phys. Rev. Lett.}{99}{2007}{066602}

\bibitem{rubinstein-prb94}
  \Name{M. Rubinstein}
  \REVIEW{Phys. REv. B}{50}{1994}{3830}

\bibitem{gu+prb96}
  \Name{R.~Y. Gu, L. Sheng, D.~Y. Xing, Z.~D. Wang, and J.~M. Dong}
  \REVIEW{Phys. Rev. B}{53}{1996}{11685}

\bibitem{pogorelov+prb98}
  \Name{Yu.~G. Pogorelov, M.~M. de~Azevedo, and J.~B. Sousa}
  \REVIEW{Phys. Rev. B}{58}{1998}{425}

\bibitem{beloborodov+rmp07}
  \Name{I.~S. Beloborodov \etal}
  \REVIEW{Rev. Mod. Phys.}{79}{2007}{469}

\bibitem{ioffe87}
  \Name{L.~B. Ioffe and V.~M. Vinokur}
  \REVIEW{J. Phys. C}{20}{1987}{6149}

\bibitem{blatter+rmp94}
  \Name{G. Blatter \etal}
  \REVIEW{Rev. Mod. Phys.}{66}{1994}{1125};
  \Name{T. Nattermann and S. Scheidl}
  \REVIEW{Adv. Phys.}{49}{2000}{607}

\bibitem{brazovskii+ap04}
  \Name{S. Brazovskii and T. Nattermann}
  \REVIEW{Adv. Phys.}{53}{2004}{177}

\bibitem{glatz+prl03}
  \Name{A. Glatz, T. Nattermann, and V. Pokrovsky}
   \REVIEW{Phys. Rev. Lett.}{90}{2003}{047201}

\bibitem{petracic+prb04}
  \Name{O.~Petracic, A.~Glatz, and W. Kleemann}
  \REVIEW{Phys. Rev. B}{70}{2004}{214432}


\bibitem{helman+prl76}
  \Name{J.~S. Helman, and B. Abeles}
  \REVIEW{Phys. Rev. Lett.}{37}{1976}{1429}

\bibitem{feigelman-jetp83}
  \Name{M.V. Feigel'man}
  \REVIEW{Sov. Phys. JETP}{58}{1983}{1076}

\bibitem{glatz-phd}
  \Name{A. Glatz}
  PhD-thesis, {\tt http://kups.ub.uni-koeln.de/ volltexte/2004/1312/pdf/diss\_ub.pdf} (2004).

\bibitem{kleemann-armr07}
  \Name{W. Kleemann}
  \REVIEW{Annu. Rev. Mater. Res.}{37}{2007}{415}

\bibitem{haldane+jp81}
  \Name{F.D.M. Haldane, and J. Villain}
  \REVIEW{J. Physique}{42}{1981}{1673}

\bibitem{nattermann+jpc83}
  \Name{T. Nattermann}
  \REVIEW{J. Phys. C}{16}{1983}{4125}

\bibitem{lemerle+prl98}
    \Name{S. Lemerle \etal}
    \REVIEW{Phys. Rev. Lett.}{80}{1998}{849}

\end{thebibliography}
\end{document}